\documentstyle[11pt,paspconf]{article}

\def\deg{\hbox{$^\circ$}}
\def\sun{\hbox{$\odot$~}}

\def\etal{{\it et al.} }
\def\ms04{MS0440$+$0204}
\def\hawaii{Hawai$'$i~}

\begin{document}

\title{X-ray Clusters at High Redshift}

\author{I.M. Gioia\altaffilmark{1}}
\affil{Istituto di Radioastronomia del CNR, 40129 Bologna, Italy, \\
       Institute for Astronomy, Honolulu, Hawaii, 96822 USA}

\altaffiltext{1}{Visiting Astronomer at CFHT,
operated by the National Research Council of Canada, le Centre
National de la Recherche Scientifique de France and the University
of \hawaii, and at the W. M. Keck Observatory, jointly
operated by the  California Institute of Technology and the
University of California}

\begin{abstract}
As the largest gravitationally bound structures known, clusters 
provide clear constraints on the formation of structure and on the
composition of the universe. Despite their extreme importance for 
cosmology the number of clusters at high redshift (z$>$0.75) is rather 
small. There are only a few X-ray emitting examples reported
and a handful of optically-selected ones.
These clusters can provide stringent constraints on theories of large
scale structure formation, if they are massive enough. 
I will review the status of these distant X-ray selected clusters.
These objects are of special importance because their X-ray emission 
and presence of gravitational arcs imply that they are massive, comparable 
to low redshift examples, and their existence is problematic for some 
theories of structure formation.

\end{abstract}



\section{Introduction}

The very existence of a massive cluster at z $>$ 0.5 is problematic
for standard CDM theories  of hierarchical structure formation
(Evrard 1989; Peebles \etal 1989).
While this problem has been recognized for some time, it has not
been taken too seriously because of the lack of conclusive evidence
of the existence of such massive systems at high redshift.

\vskip 10pt

Naturally the statistics on the abundance of high z clusters are
poor because the number of such clusters known at present is small.
Furthermore, one might argue that many of these clusters may not be
as massive as they appear due to inflation of the measured velocity
dispersion from field galaxy contamination and projection effects
(see among others Frenk \etal 1990).
However, the number of distant clusters is steadily increasing
due to the combined use of telescopes such as HST (Hubble Space Telescope)
and Keck, as we have heard from many contributors at this meeting.

\vskip 10pt

X-ray observations of distant (z $>$ 0.7) optically selected 
clusters have shown that only a few clusters are detected 
(see among others Castander \etal 1994), and those few are relatively 
modest  X-ray emitters (L$_{x}\sim 10^{44}$ erg s$^{-1}$), with L$_{x}$ much 
lower than that of present day clusters. While this fact would 
seem to imply that extremely rich systems with large amounts of 
X-ray emitting gas do not exist at these high redshifts, I will
show several examples that indicate that this is not the case.
In Table 1, five examples of EMSS (Einstein Medium Sensitivity
Survey; Gioia \etal 1990, Stocke \etal 1991) 
clusters with average z = 0.66, and z$_{max}$ = 0.826 are listed. In 
addition to their high X-ray luminosity, there is other compelling evidence 
that these clusters are genuinely massive. Three of the z$\sim$0.6-0.7 
clusters contain large lensed arcs which allow a crude estimate 
of the projected mass in the cluster cores. For four of them weak 
lensing studies have been performed. 
The letter ``s'' in the third column of Table 1 indicates the 
presence of giant arcs in the cluster core, while the letter 
``w'' indicates that weak lensing analyses have been performed. 
L$_{x}$ is given in units of 10$^{44}h^{-2}_{50}$ erg s$^{-1}$
in the 0.3$-$3.5 keV band, T$_{x}$ is measured in the 2$-$10 keV band
and N$_{0.5}$ is the central richness of the cluster
(defined in Bahcall 1981; notice that
in the same units Coma has N$_{0.5}$=28).

\vskip 10pt

\begin{table}
\caption{EMSS Clusters with z $>$ 0.5} \label{tbl-1}
\begin{center}\scriptsize
\begin{tabular}{lrrrrll}
Name & redshift & lensing & L$_{x}$ (erg s$^{-1}$)& T$_{x}$ (keV)& N$_{0.5}$ & $\sigma$ (km s$^{-1}$)\\
\tableline
MS0015.9$+$1609&0.546\tablenotemark{a}& w &14.64\tablenotemark{b}&8.4\tablenotemark{c}&66$\pm6$\tablenotemark{d}&1324\tablenotemark{a}\\
MS0451.6$-$0305&0.550\tablenotemark{e}& sw&19.98\tablenotemark{b}&10.4\tablenotemark{f}&47$\pm5$\tablenotemark{d}&1371\tablenotemark{g}\\
MS1054.5$-$0321&0.826\tablenotemark{b}& w & 9.28\tablenotemark{b}&14.7\tablenotemark{h}&82$\pm10$\tablenotemark{i}&1360\tablenotemark{h}\\
MS1137.5$+$6625&0.782\tablenotemark{b}& sw& 7.56\tablenotemark{b}&    &56$\pm6$\tablenotemark{d}&\\
MS2053.7$-$0440&0.583\tablenotemark{e}&  s& 5.78\tablenotemark{b}&   &20$\pm5$\tablenotemark{d} &\\

\end{tabular}
\end{center}
\tablenotetext{a}{Dressler \& Gunn 1992}
\tablenotetext{b}{Gioia \& Luppino 1994}
\tablenotetext{c}{Furuzawa \etal 1994}
\tablenotetext{d}{Luppino \& Gioia 1995}
\tablenotetext{e}{Maccacaro \etal 1994}
\tablenotetext{f}{Donahue 1996}
\tablenotetext{g}{Carlberg \etal 1996}
\tablenotetext{h}{Donahue \etal 1997}
\tablenotetext{i}{Luppino \& Kaiser 1997}

\end{table}

\section{EMSS z $>$ 0.5 clusters}

Since its discovery MS0015.9$+$16 (alias CL0016$+$16, Koo 1981)
has been the archetypal rich, X-ray luminous, distant cluster. It 
is part of the EMSS since it was ``rediscovered'' in an 
observation pointed at another target. The cluster is very rich
and has a linear structure elongated in the NE-SW direction.
Smail \etal  (1994) reported detection of weak shear,
which has been reconfirmed by the analysis performed by Luppino \etal
(1996) using different deep images with a larger field of view
and a different technique (Kaiser and Squires, 1993; Squires and Kaiser,
1996). Two hours of R band  Keck data are in hand but the images have 
not been assembled yet (Clowe, private communication).

\vskip 10pt

MS0451$-$03 with an L$_{x}$ of almost 2$\times$10$^{45}$erg s$^{-1}$
is the most X-ray luminous cluster in the EMSS and among the
brightest clusters known. ASCA data by Donahue (1996) show a hot cluster at
10.4 Kev with iron abundance of 15\% solar and a total mass
(within 1h$_{50}^{-1} $Mpc) of 9.7$_{-2.2}^{+3.8}\times10^{14}$ M\sun.
The shear signal
detection, performed by Luppino \etal (1996) on a 7200s R band 
image taken with the UH 2.2m telescope is at $>$5$\sigma$. Two hours of R 
band  Keck data are in hand also for this cluster but the
analysis has not been performed yet.

\vskip 10pt

MS1054$-$03 is the most distant cluster of the EMSS and among the most
distant X-ray selected clusters known. The cluster has a 
filamentary morphology with the X-ray coming from the center and
elongated in the same E$-$W direction as the optical galaxies
(Donahue \etal 1997).  MS1054$-$03 is extremely rich and quite
hot at 14.7$^{+4.6}_{-3.5}$ keV as obtained by ASCA (Donahue \etal 1997).
A strong shear signal at 6$\sigma$ level is detected
(Luppino \& Kaiser, 1997). The total mass (within 1 Mpc) from X-rays 
and from weak lensing are consistent
(2-6$\times 10^{14} h_{50}^{-1}$ M\sun vs 3-30$\times 10^{14} h_{50}^{-1}$
M\sun). 

\vskip 10pt

Deep imaging of MS1137$+$66 with Keck and with the UH 2.2m have been
collected. From the reduced images a large arc has been
discovered close to the cluster center (Clowe \etal 1997). 
Differently from MS1054$-$03 and MS0015.9
$+$16 this cluster is compact and concentrated with no filamentary
structure. A weak lensing analysis, performed by
Clowe \etal  on a 8700 second R-band exposure of Keck, using 
the I band (7500 s) 2.2m data as a color selection to remove cluster 
galaxies, finds a nice centrally concentrated mass peak falling 
exactly on the brightest cluster galaxy. The mass from 
weak lensing comes out to be 2.9$\times 10^{14} h^{-1}$ M\sun 
at 500 $h^{-1}$ kpc (assuming the background galaxies lie at z=2).

\vskip 10pt

Also for MS2053-04 two hours of R band  Keck data are in hand but the
images have not been assembled. A recent observation of
this cluster has been performed with the Italian-Dutch BeppoSAX
satellite by Scaramella \etal (in preparation). The data
have not been fully reduced yet. A preliminary analysis shows this 
cluster to be much cooler than MS1054$-$03 or MS0451$-$03.

\section{The ROSAT NEP Survey}

The  North Ecliptic Pole (NEP) region of the Rosat All-Sky Survey 
(RASS; Tr\"umper 1991) has the largest exposure time
(approaching 10 ks) of the all RASS. The NEP
region covers a 9$\deg \times 9\deg$ field, and contains a total of
465 X-ray sources detected at $>$ 4$\sigma$ in the 0.1$-$2.4 keV
(Mullis \etal 1998). We are identifying all sources in the field. 
The principal derivative is a statistically complete sample of 
galaxy clusters appropriate for a better 
characterization of the X-ray luminosity function evolution.
We have discovered a very distant cluster in the NEP 
at z= 0.81. RXJ1716$+$66 (Henry \etal 1997) is among the most distant 
X-ray selected  clusters together with MS1054$-$03 and the X-ray 
clusters detected by Rosati (1997) in the RDCS (ROSAT Deep Cluster 
Survey) or Ebeling \etal (1997) in the WARPS (Wide Angle 
ROSAT Pointed Survey).

\subsection{RXJ1716$+$66}

As with MS1054$-$03, it is not likely that  RXJ1716$+$66
is in virial equilibrium.  The galaxies in RXJ1716$+$66 are 
in an inverted S-shaped filament running northeast to southwest 
(Fig. 1). Cluster members extend all along the S. The 
distance from the top to the bottom of the S is about 1.5 Mpc 
(H$_{0}=$50 km s$^{-1}$ Mpc$^{-1}$, q$_{0}$=0.5).
X-ray follow-up observations with ASCA (100 ks) and  with the 
ROSAT HRI (171 ks) provide a temperature in 2$-$10 kev 
of kT = 6.7 $^{+3.2}_{-1.8}$ (68\% confidence) and a flux in
0.5$-$2.0 keV of 1.16$\pm$0.33 $\times$ 10$^{-13}$  
erg cm$^{-2}$s$^{-1}$. It is intriguing that the morphology of
RXJ1716$+$66 and MS1054$-$03 is filamentary with the X-rays coming
from the center. We note this since the initial formation of
protoclusters is often described as matter flowing along filaments
(Bond \etal 1996) with the X-rays generated at the impact point of the
two colliding streams of matter (Henry \etal 1997).

\vskip 11.5truecm
{\small
Fig. 1-A 1024$\times$1024 subarray image of RXJ1716$+$66 extracted from
the center of a 4500s exposure in the I-band taken by G. Luppino
and M. Metzger with the UH 8K$\times$8K CCD mosaic camera on the CFHT
prime focus in Sept 1995. North is up and East to the left. This image
spans 3.\arcmin6$\times$3.\arcmin6 (0.9h$^{-1}$ Mpc at $z=$0.81)
at a scale of 0.\arcsec21/pixel.}

\vskip 10pt

Both CFHT 8K$\times$8K mosaic CCD deep images 
and Keck R-band (7500s) images were taken. The
weak lensing analysis was performed on the R-band Keck image, using 
2.2m I-band data (26,100s) for color terms (Clowe \etal 1997). 
The mass peak is just east of the BCG, and an arm of mass to the 
NE, neatly following  the line of galaxies NE of the cluster core
is detected. Doug Clowe' analysis gives a very strong signal of 
4$\times 10^{14} h_{-1}$ M\sun (background galaxies at z=1.5).
The M/L$_{V}$ is equal to 210$h$. Both the optical lightmap and 
weak lensing massmap have two spatially distinct massive 
sub-clusters, as well as a long filamentary structure.

\section{Conclusions}

There is clear evidence from the existing data 
that large, dense mass concentrations existed
at an early epoch. Another possible search strategy to 
find distant X-ray clusters is to look for clusters around 
powerful radio galaxies. High z massive structures have been found 
around several 3C radio galaxies (i.e.: 3C184, at z$=$0.99, 
Deltorn \etal 1997; 3C324, at z$=$1.2, Smail \& Dickinson, 1995,
just to quote  a few). Cluster abundances has been cited as 
one of the strongest evidence against the standard CDM model as
normalized to reproduce the microwave background anisotropies 
seen by COBE satellite. The number of high-z massive clusters 
predicted by standard CDM, or other mixed Dark Matter models, is
too low with respect to the number of clusters observed. In low-density
universes there is less evolution and nearly all 10$^{15}$ M\sun
clusters formed by z$=$0.5. In a high-density universe, instead,
only 5\% of the present day 10$^{15}$ M\sun clusters
have formed by the same redshift. Thus massive clusters should be 
much rarer at epochs earlier than 0.5 if $\Omega$=1, contrary to the
observations. Alternatives have been suggested with low $\Omega$
models (Viana \& Liddle, 1996; Eke \etal 1996).
The open CDM ($\Omega$=0.3) and $\Lambda$-dominated CDM models
are preferred because they are compatible with the data.
It may be possible to estimate $\Omega$ from forthcoming observations 
of intermediate-distant clusters (up to z$\sim$1). 

\acknowledgments

I am grateful to P. Henry, C. Mullis, N. Kaiser, G. Luppino, 
M. Donahue, D. Clowe  for stimulation, help
and advice. This work has received partial finacial support from
NASA-STScI grant GO-5402.01-93A and GO-05987.02-94A,
from NSF AST95-00515, from NASA grant NAG5-2594
and ASI grants ARS-94-10 and ARS-96-13.

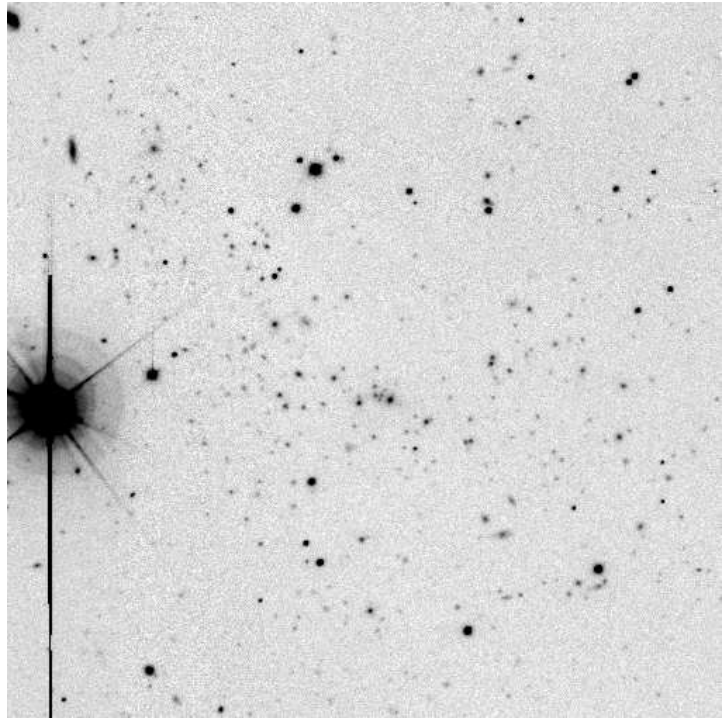
\end{document}